\documentclass[useAMS]{mn2e}

\usepackage{graphicx}
\usepackage{times}
\title[M-type Mira variables of moderately different mass]
{Pulsation of M-type Mira variables with moderately different mass:
search for observable mass effects}

\author[M.J. Ireand et al.]{M.J. Ireland$^1$, M. Scholz$^{1,2}$, 
P.G. Tuthill$^1$, P.R. Wood$^3$\\
$^1$School of Physics, University of Sydney NSW 2006, Australia\\
$^2$Institut f\"{u}r Theoretische Astrophysik der Universit\"{a}t Heidelberg,
Albert-\"{U}berle-Str.2, 69120 Heidelberg, Germany\\
$^3$Research School for Astronomy and Astrophysics, Australian National 
University, Canberra ACT 2600, Australia}

\begin{document}

\pagerange{\pageref{firstpage}--\pageref{lastpage}} \pubyear{2004}

\maketitle

\label{firstpage}

\begin{abstract}
Models of M-type Miras with masses of 1 $M_\odot$ and 1.2 $M_\odot$, i.e.
with envelope masses of about 0.4 $M_\odot$ and 0.6 $M_\odot$, have been
constructed, and a comparison has been made of their observable properties.
Geometric pulsation of continuum-forming layers is found to be little affected 
by the mass difference. 
The influence of molecular contamination of near-infrared continuum bandpasses 
upon interferometrically measured fit diameters ranges from undetectable to 
quite significant. 
Some pulsation cycles of the lower-mass model Mira show substantially stronger 
contamination than that found in any cycle of the higher-mass star.
Observations which sample pulsation phase well and continuously are crucial for 
avoiding misinterpretations, because the assignment of absolute pulsation 
phases is inherently uncertain by at least 0.1 cycles, diameter changes may be 
strongly phase-dependent, and cycle-to-cycle variations may be substantial. 
In accord with expectations, we find that cycle-to-cycle variations that show 
up in light curves and in near-continuum diameters tend to be larger and more 
common in the low-mass models, leading to one possible way to discriminate mass.
Two other methods, based on high-precision measurements of the pulsation
amplitude and on derivation of pre-maximum effective temperatures from 
diameter measurements, are also discussed.
High-layer features that
may be strongly affected by mass are not well described by present dust-free
models.
\end{abstract}


\section{Introduction}

The typical mass of an M-type Mira variable is assumed to be of the order of
1 $M_\odot$ from general considerations of AGB evolution and pulsation theory.
This value is uncertain by a few tenths of a solar mass, but no
method is known for accurately determining the mass of a non-binary Mira. 
Wyatt \& Cahn (1983) derive masses between 1.00 and 1.66 $M_{\odot}$ for 
the progenitor main-sequence stars of 124 Miras (10 percent of these larger 
than 1.30 $M_{\odot}$) from an analysis of their kinematic properties.
The "round" value of 1.0 $M_\odot$ is often adopted in Mira modelling. Recent 
model studies of the disk brightness distribution of these stars, however, seem
to indicate that moderate changes of the stellar mass lead to effects on 
geometric pulsation (Hofmann, Scholz \& Wood 1998, henceforth HSW98; 
Jacob \& Scholz 2002, henceforth JS02) that
might be accessible to properly designed observations. From the uniform-disk
(UD, or other simple disk-brightness) fit radii given by HSW98 and JS02, one 
suspects that both the pulsation of continuum-forming layers and the molecular 
contamination of near-continuum bandpasses that partly masks geometric 
pulsation (Ireland, Scholz \& Wood 2004, henceforth ISW04) may depend on mass. 
Since, however, the HSW98 and JS02 studies consider only two phases per
cycle and centre-to-limb variation (CLV) of the disk intensity often varies 
strongly with phase (ISW04), a more elaborated model set is needed for
assessing mass effects. We compare in this study two series of Mira pulsation 
models which have identical luminosity and linear period 
but a moderately different mass for the parent star, and look for observable 
pulsation differences.

There are only a few interferometric observation of the periodic variation 
of the near-continuum angular diameter available in the literature:
Tuthill, Haniff \& Baldwin (1995), $o$~Cet, 0.833~$\mu$m, 0.902~$\mu$m; 
Burns et al. (1998), R~Leo, 0.833~$\mu$m, 0.940~$\mu$m; 
Perrin et al. (1999), R~Leo, 2.2~$\mu$m K bandpass; 
Young et al. (2000), $\chi$~Cyg, S-type Mira, 0.905~$\mu$m; 
Thompson, Creech-Eakman \& Van Belle (2002), S~Lac, 2.2~$\mu$m K bandpass;  
Woodruff et al. (2004), $o$~Cet, 2.2~$\mu$m K bandpass. 
None of these observations covers one or more full cycles, and observed
phase intervals tend to be fairly narrow. They probe a combination of 
continuum pulsation and phase-dependent molecular-band absorption in the 
observed bandpasses (ISW04).

\section{Models}

\begin{figure}
\hspace{-0.3 cm}
\includegraphics[scale=0.47]{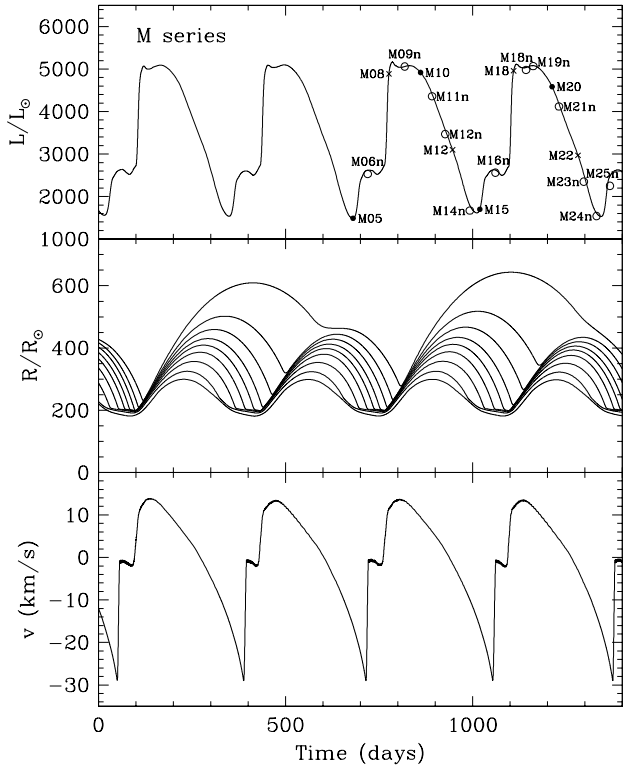}
 \caption{Luminosity (top panel), radii of selected mass zones (middle) and 
  the velocity of the 5th mass zone from the centre (bottom) of the M model
  series (see HSW98). One cycle is covered by about 1400 pulsation model 
time-steps. Those time-steps for which a detailed non-gray atmospheric model was
constructed (see HSW98) and whose properties are discussed in this study are
marked by dots (models of HSW98), crosses (TLSW03) and circles (this work).}
 \label{figPhaseFig}
\end{figure}

\begin{figure}
\hspace{-0.5 cm}
\includegraphics[scale=0.47]{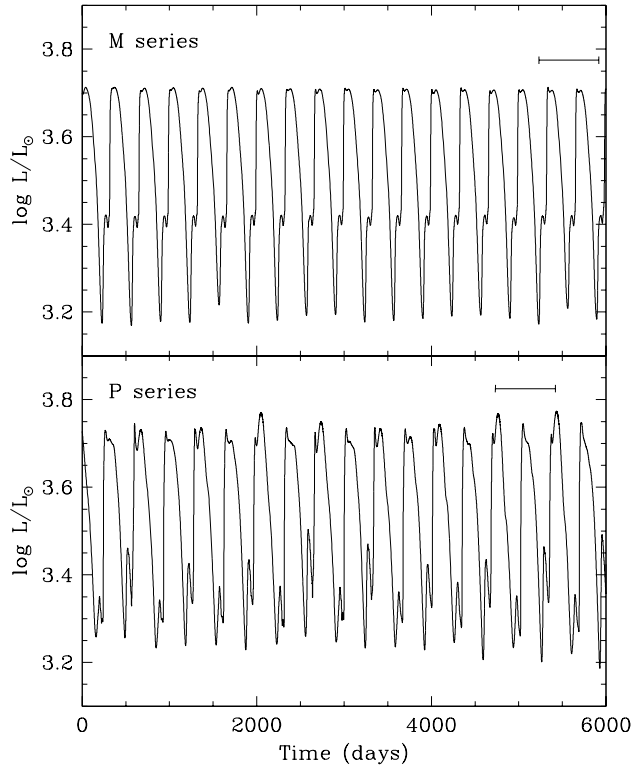}
 \caption{Luminosity for many cycles of the M model series (top panel) and
   P model series (lower panel). The horizontal bars indicate the range of 
   phases used in this paper.}
 \label{figRegularity}
\end{figure}

This study is based upon two series of complete self-excited models of HSW98,
supplemented by additional phases given in Tej et al. (2003b, henceforth 
TLSW03), in ISW04 and in this paper. The parameters of these series were chosen
to represent the M-type Mira prototypes $o$~Cet and R~Leo. The non-pulsating
parent star has solar metallicity, luminosity $L/L_\odot$=3470, and mass
$M/M_\odot$=1.0 for the P model series and $M/M_\odot$=1.2 for the M series.
The linear pulsation period is 332 days in both cases. This comparative
study is aimed at seeing whether, at a given period, observable characteristics
vary in a way that might be used to estimate the star's mass. Note that the 
20 percent increase in mass does, in fact, imply an increase of about 
50 percent in envelope mass as the typical AGB core mass is about 0.6 solar 
masses. The $\tau_{\rm Ross}$=1 Rosseland radius of the parent star is 
$R_{\rm p}/R_\odot$=241 for the P series and $R_{\rm p}/R_\odot$=260 for the 
M series, corresponding to an effective temperature 
$T_{\rm eff} \propto (L/R^2)^{1/4}$ of 2860K (P) and 2750K (M). 

In the HSW98 model series, the mixing length $l$ enters the pulsation 
calculation as a free parameter that is so chosen that, for the adopted 
luminosity and mass, the pulsation period was close to the 332 day period of 
$o$~Cet. It is 2.06$H_p$ for the P models and 1.73$H_p$ for the M models 
($H_p$ = pressure scale height). Since, however, the choice of $l$ is only a 
very minor factor in overall pulsation behaviour, predicted differences between
the two model series are essentially generated by the difference of mass.
Details regarding assumptions of modelling pulsation and the dynamic 
atmosphere are described in HSW98. The models of TLSW03 and ISW04 and the new 
models given here are constructed in the same way as the HSW98 models.

Using the LMC period-luminosity relation (e.g. Feast et al. 1989;
Hughes \& Wood 1990) the adopted luminosity of the model series
appears to be significantly too low. However, the distances required
for the model series to match the J- and K-band magnitudes of the
prototype Mira stars $o$~Cet and R~Leo is within $2\sigma$ of their
HIPPARCOS distances given by Knapp et al. (2003). The HIPPARCOS
distances would seem to indicate that the adopted luminosity is
slightly too high for $o$~Cet and slightly too low for R~Leo,
resulting in an unclear answer for the best model luminosity to adopt
(see the discussion in ISW04).

Since the P model series of HSW98 have cycles with strikingly different
amounts of molecular contamination of near-continuum bandpasses (JS02), ISW04
selected 2 successive cycles (of 4 cycles given in HSW98) that show a very low 
and a very high contamination. Inspection of a large sequence of cycles of our
two model series shows that cycle-to-cycle variations are still present but 
much less pronounced in the M series, and the 2 cycles given in HSW98 and in 
this paper should illustrate the typically very modest differences between 
cycles in a more massive Mira. Since phase assignment is only approximate for 
the models listed in the HSW98/TLSW03 tables, we have re-assigned more 
accurate phases to the full set of HSW98/TLSW03 models and our new set of 
supplementary models in order to study detailed phase effects. The visual 
phases which are used to identify the models of the series are approximate 
and precede bolometric phases by 0.1 (Lockwood \& Wing 1971; cf. HSW98). 
Note, however, that the absolute zero point of any phase assignment is 
uncertain by at least 0.05 to 0.1 due to the irregularities of modelled 
(HSW98) and observed (e.g. Whitelock, Marang \& Feast 2000) Mira light 
curves. Furthermore, the actual pulsation period that arises after pulsation 
sets in differs slightly from the linear pulsation period of the parent 
star. It is about 317 days for the P model series and about 341 days for the 
M series. Individual cycle lengths may readily vary by a few days. 
Altogether, absolute phases in Table 1 are uncertain by at least 0.1 
whereas relative phases are accurate to about 0.01 to 0.02.

Figure~\ref{figPhaseFig} is the equivalent of Figure 2 of HSW98 showing 
the phase positions of all M models of HSW98, TLSW03 and this paper. Its
counterpart for the P models is plotted in ISW04. Note that the outer
layers of the model as seen in the central panel do not show the same
periodicity as the central star, and that by examining only two full
successive cycles we can include much of the cycle-to-cycle variation
that one might observe using a longer time series (this is also clear
in the P series plot in ISW04). Figure~\ref{figRegularity} shows a much
longer time series of luminosity for both the P and M series models,
showing the positions of the model series examined in this paper. It
is clear from this plot that the M series has significantly less
cycle-to-cycle variation than the P series.
It is a general feature of extant red giant pulsation models that 
reducing the envelope mass increases the irregularity of the pulsation. 
The increased irregularity at lower total mass
is probably related to the increasingly chaotic
behaviour of the outer layers as the envelope mass is decreased (Icke,
Franck and Heske 1992).
Here, the M model envelope mass (about 
0.6 $M_\odot$) is 50 percent higher than that of the P series (0.4 $M_\odot$).
Note that the cycles selected for detailed investigation in this paper are 
not particularly unusual.

The full set of M models is given in Table 1. In addition to the standard 
$\tau_{\rm Ross}$=1 Rosseland radius $R$ and the corresponding effective 
temperature $T_{\rm eff} \propto (L/R^2)^{1/4}$, Table 1 also gives 
the $R_{1.04}$ filter radius, defined after Scholz \& Takeda (1987) in 
a narrow 1.04~$\mu$m bandpass (centred at 1.0465~$\mu$m, width
0.001~$\mu$m, see JS02) and the corresponding effective 
temperature $T_{1.04} \propto (L/R_{1.04}^2)^{1/4}$. Since this bandpass shows 
very little contamination, $R_{1.04}$ is essentially the monochromatic 
$\tau_{1.04}$=1 optical-depth radius and describes the approximate position
of the continuum-forming layers. Though this is a non-observable quantity,
the ISW04 study shows that it is very close to an observable interferometric 
fit radius and probably is an excellent indicator of geometric pulsation 
of an M-type Mira variable. Note that the position of the $\tau_{\rm Ross}$=1 
layer is not well suited for describing geometric pulsation owing to the 
properties of the Rosseland mean of extinction coefficients (see e.g. HSW98; 
Scholz 2003; ISW04).

For both the P and the M model series, the transition region between
the dust-free stellar atmosphere and the dust-driven stellar wind region is a
significant cause of modelling uncertainty. These model series do not
include dust formation. They have an arbitrary radius cutoff of 5 $R_p$, and
a minimum temperature of 740~K is adopted in the outermost layers where the 
state equation of gas particles is assumed to become unrealistic. Dust could 
form from about 2 to 3 continuum radii (e.g. Danchi et al. 1994; 
Danchi \& Bester 1995; Habing 1996; Lobel et al. 2000; 
Lorenz-Martins \& Pompeia 2000). It would influence the density stratification
of the outer part of the atmosphere due to radiation pressure as well as the 
temperature stratification due to radiation transport effects in case 
of noticeable thermodynamic coupling of dust and gas particles (cf. 
Bedding et al. 2001). In the dust-free M models, the temperature of about 
1200 K where silicate grains may be formed is reached around 2 continuum radii 
(e.g. defined as $R_{1.04}$) at most phases
and at about 3 continuum radii at pre-maximum phases of our 2nd cycle.
In the dust-free P models, the 1200 K distance is more phase- and
cycle-dependent and ranges from around 2 to more than 4 continuum radii.
Corundum condensation occurs at about 1400 K but may be insignificant in 
M-type atmospheres (e.g. Gail \& Sedlmayr 1998; cf. also Patzer 2004).
Note, however, that dust temperatures in a dusty atmosphere tend to be higher 
at a given distance from the star's centre than gas temperatures in a
dust-free 
atmosphere, with details depending upon thermodynamic coupling between dust
and gas (Bedding et al. 2001, e.g. their Figures 1 and 2 showing the models
P20 and M10). Therefore, comparisons between the model series 
in this paper will focus on the wavelength regimes that are relatively
insensitive to these outer atmospheric layers (i.e. where there is a
relatively small optical depth at 2 to 3 continuum radii). 

\begin{table}
 \caption{Parameters of M series Mira models. 
  The model names ending in `n' are the new models presented in this paper. The
  columns: visual phase $\phi_{vis}$; luminosity $L$; Rosseland radius
  $R$; 1.04~$\mu$m near-continuum radius $R_{1.04}$; effective
  temperatures $T_{\rm eff}$ and $T_{1.04}$ corresponding to $R$ and 
  $R_{1.04}$; reference for previously published models.}
\begin{tabular}{@{}llllllll@{}}
 \hline
  Mod.  & $\phi_{vis}$ & $L$ & $R$ & $R_{1.04}$ & $T_{eff}$ & $T_{1.04}$ & Ref. \\
        &     & ($L_{\odot}$)&($R_p$) &($R_p$)  & (K)       & (K)        &\\ 
 \hline 
M05  & 0+0.49& 1470 & 0.93 & 0.84 & 2310 & 2420 & HSW98\\ 
M06n & 0+0.60& 2430 & 0.78 & 0.78 & 2860 & 2860 &  \\
M08  & 0+0.77& 4780 & 0.81 & 0.81 & 3320 & 3320 & TLSW03\\
M09n & 0+0.89& 5060 & 1.03 & 1.03 & 2970 & 2970 &  \\
M10  & 1+0.02& 4910 & 1.19 & 1.18 & 2750 & 2760 & HSW98\\
M11n & 1+0.11& 4360 & 1.26 & 1.21 & 2590 & 2640 &  \\
M12n & 1+0.21& 3470 & 1.30 & 1.18 & 2410 & 2540 &  \\
M12  & 1+0.27& 2990 & 1.33 & 1.12 & 2300 & 2500 & TLSW03 \\
M14n & 1+0.40& 1670 & 1.17 & 0.91 & 2110 & 2400 &  \\
M15  & 1+0.48& 1720 & 0.88 & 0.83 & 2460 & 2530 & HSW98\\
M16n & 1+0.60& 2460 & 0.77 & 0.77 & 2860 & 2860 &  \\
M18  & 1+0.75& 4840 & 0.81 & 0.81 & 3310 & 3310 & TLSW03\\
M18n & 1+0.84& 4980 & 0.99 & 1.00 & 3020 & 3010 &  \\
M19n & 1+0.90& 5070 & 1.08 & 1.09 & 2900 & 2900 &  \\
M20  & 2+0.05& 4550 & 1.23 & 1.20 & 2650 & 2680 & HSW98\\
M21n & 2+0.10& 4120 & 1.26 & 1.21 & 2550 & 2610 &  \\
M22  & 2+0.25& 2850 & 1.27 & 1.10 & 2330 & 2490 & TLSW03\\
M23n & 2+0.30& 2350 & 1.25 & 1.03 & 2230 & 2460 &  \\
M24n & 2+0.40& 1540 & 1.09 & 0.87 & 2160 & 2410 &  \\
M25n & 2+0.50& 2250 & 0.80 & 0.79 & 2770 & 2780 &  \\

 \hline
\end{tabular}
 \label{tblModelParams}
\end{table}

\section{Light curves}

\begin{figure*}
 \includegraphics{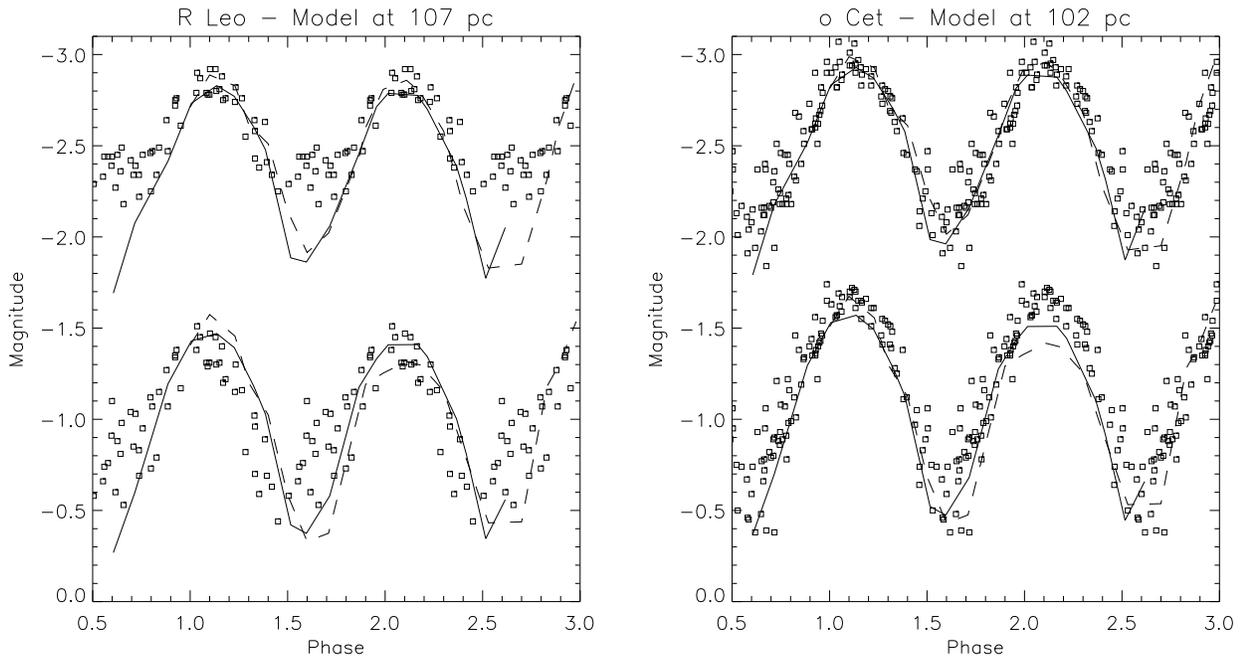}
 \caption{Light curves for the P (dashed line) and M (solid line) model
  series (K-band - upper plot, 
  J-band - lower plot) with observations of Whitelock et al. (2000)
  overplotted. See text for phase adjustment of M series. }
 \label{figWhitelockComp}
\end{figure*}


\begin{figure*}
 \includegraphics{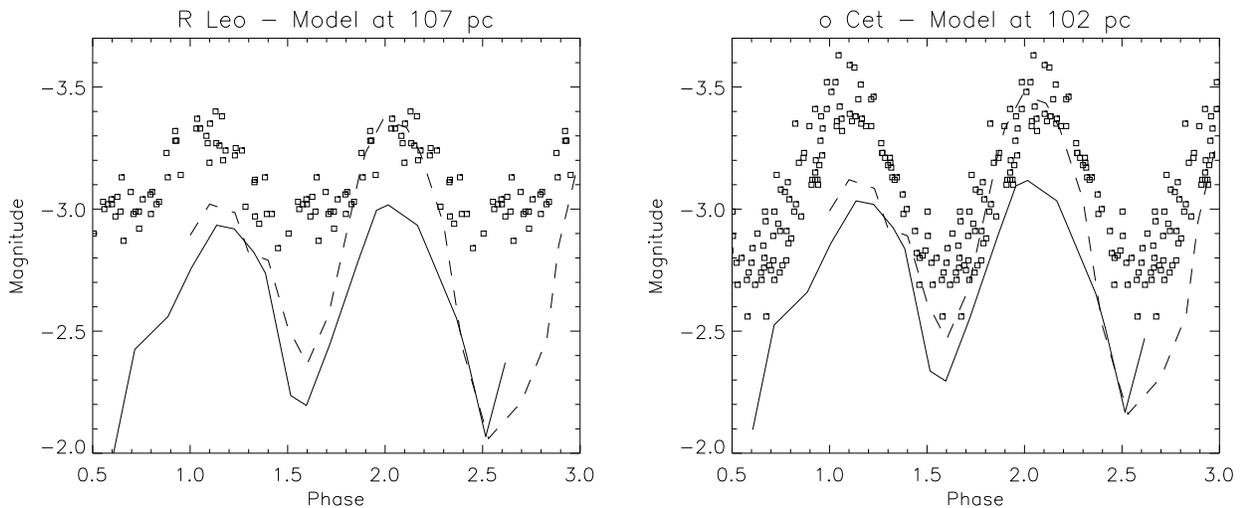}
 \caption{Light curves for the P (dashed line) and M (solid line) model
  series for the L band (see text) with observations of Whitelock et
  al. (2000) overplotted.}
 \label{figLBand}
\end{figure*}

The model-predicted light curves in the K (simple rectangular
  bandpass, centred at 2.195~$\mu$m, width 0.4~$\mu$m, see JS02) and
  J  (1.25,0.3) bandpasses may be 
compared with the observations of $o$~Cet and R~Leo by Whitelock et al. (2000)
which cover a large number of pulsation cycles and show clearly significant
cycle-to-cycle variations.
Figure~\ref{figWhitelockComp} 
is a fit of model-predicted light curves covering two successive 
cycles of the P series (from model P10 at phase 1.00 to P30 at 2.98, Table 1 of
ISW04) as well as the M series (from M05 at 0.49 to M25 at 2.50, Table 1) to
the observations. 

The fit includes a phase zero-point shift as well as a magnitude zero-point
shift in the K bandpass (other filters are then fixed by this zero point)
so that good 
agreement occurs near maximum. The visual minus K-band phase shift of M-type 
Miras is just marginally larger than the visual minus bolometric phase shift of
0.1 adopted in model phase assignment (difference less than 0.05, 
Smith et al. 2002). The difference between the 
K-fit phase shifts of our two model series (0.116) will be used throughout
this paper for comparing the phase dependence of model-predicted quantities, 
i.e. +0.116 is added to M model phases of Table 1 in Figure 2 and all 
forthcoming figures.
The model stars of both series would have to be placed at distances of 
107~pc and 102~pc for emitting the near-maximum K magnitudes quoted by
Whitelock et al. (2000) for the stars $o$~Cet and R~Leo respectively
(ISW04, see the discussion given there as for HIPPARCOS distances). 
Note also that the periods 
of $o$~Cet (334 days) and R~Leo (313 days) given by Whitelock et al. (2000) 
differ slightly from the model periods discussed in Section 2 (linear
pulsation period of parent star: 332 days; period of P series: 341 days; 
period of M series: 317 days).

The agreement between model-predicted and observed light curves
(Figure~\ref{figWhitelockComp})
is good, though neither star is perfectly modelled by the P or the M series
and some significant differences are quite obvious, in particular the smaller 
amplitude of R~Leo in the K bandpass. 
In fact, an excellent fit by the M model curve may be achieved for 
$o$~Cet in both the K and J bandpasses if the M model star were placed at a 
slightly smaller distance, i.e. if the M curve were slightly shifted upwards in
the figure. No comparably good fit by the P model curve is possible because of 
the much larger difference between the two cycles. The small K-band amplitude 
of R~Leo is not predicted by either model series. Note that differences between
our simple rectangular bandpass profiles and those used in the observations of 
Whitelock et al. (2000) account for only very small effects (see ISW04).
We did not attempt to fit optical light curves as these are much
less reliably predicted by the models owing to excessive temperature 
sensitivity in the Wien part of Planck functions and to the uncertainties 
of treatment of very strong TiO bands (cf. TLSW03).

Figure~\ref{figLBand} shows the model light curves for the
non-rectangular L bandpass as defined in Bessell and Brett (1988) with
observations of Whitelock et al. (2000) overplotted. The cycle-to
cycle variations of the P series are clearly much more pronounced than
either the M series or the observations. Furthermore, both model
series (with the exception of the second maximum of the P series)
have systematically lower fluxes than either star. 
This discrepancy is not surprising, because this bandpass 
(about 3.2-3.75~$\mu$m) is noticeably affected by relatively strong
H$_2$O absorption in the outer layers of the model star 
(e.g. Tej, Lan\c{c}on \& Scholz 2003a), where there are 
significant modelling uncertainties (see Section 2). In contrast, bandpasses
that are conventionally used as near-continuum bandpasses in interferometric
work (e.g. Mennesson et al. 2002, L' 3.4-4.1; JS02, 
L* 3.5-4.1) avoid the strong-H$_2$O regime at the short-$\lambda$ end of the
L filter and are not so extremely sensitive to the structure of the outer 
atmosphere. Thus, reliable information about mass effects may be predicted
by presently available models only from bandpasses with relatively little 
contamination from the outermost low-temperature layers of the star.
It should be noted in this context that studies based upon empirical H$_2$O 
shells that may reproduce contamination effects in near-continuum bandpasses 
quite satisfactorily (e.g. Mennesson et al. 2002; Ohnaka 2004; Weiner 2004) 
are unable to predict the time and parameter dependent behavior of molecular 
absorption in a real Mira star.

We have also made a close comparison between the model photometry of
the P and M series in near-continuum bandpasses, in order to search
for mass indicators based on colour. We find that the
cycle-to-cycle variation masks any possible systematic differences
in colours between the two series, despite 
the fact that the M models are systematically cooler than the
P models at near-maximum phases (cf. Table 1 with the table of ISW04).

\section{Interferometric Diameters}

\begin{figure}
 \includegraphics{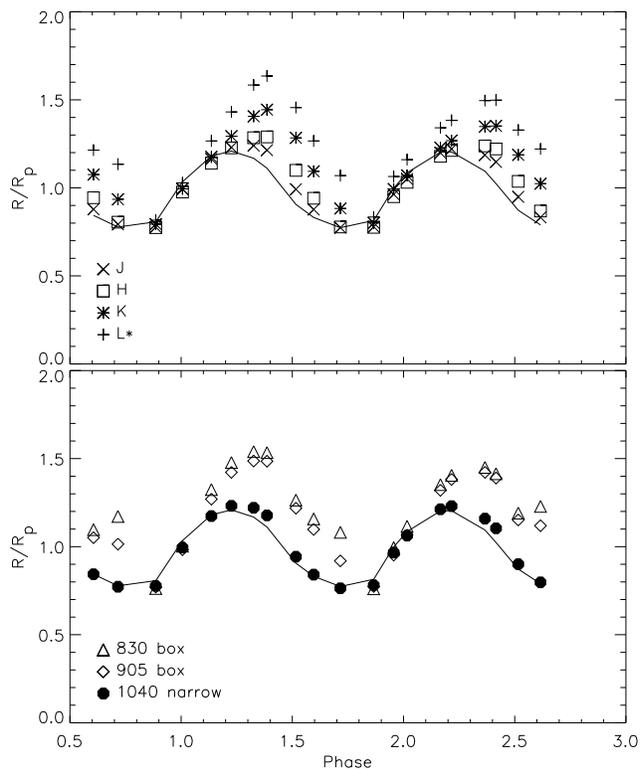}
 \caption{Radii from single-point V=0.3 uniform disk fits to the M model
  series in 7 near-continuum bandpasses (see text). The overplotted
  solid line in both panels is the $\tau$=1 optical-depth radii for
  the 1.04~$\mu$m continuum bandpass ($R_{1.04}$).}
 \label{figFitDiams}
\end{figure}

\begin{figure}
 \includegraphics{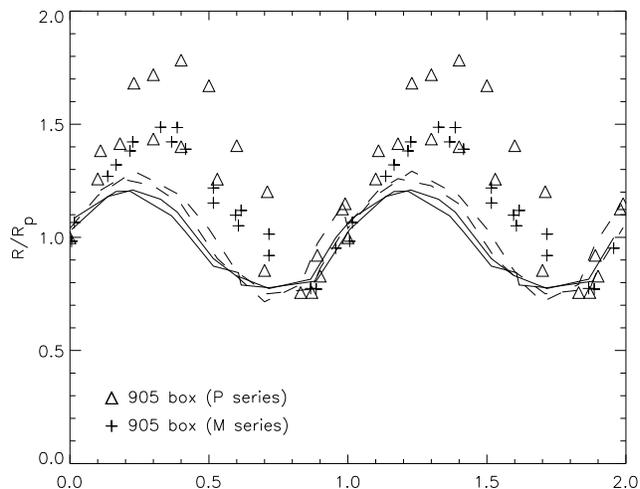}
 \caption{Radii from single-point V=0.3 uniform disk fits to both the
  P and M model series in the 905~nm bandpass, which is moderately
  contaminated by the TiO molecule. The $R_{1.04}$ radii (see text)
  are overplotted for the M (solid line) and P (dashed line) series.
  Both cycles of both series are folded-in to a single cycle 
  and repeated so that comparison of any systematic differences
  between the models can be made.}
 \label{fig905Diams}
\end{figure}

\begin{figure}
 \includegraphics{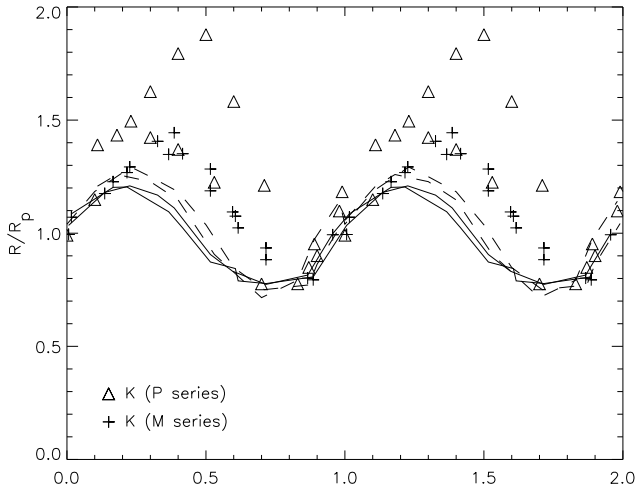}
 \caption{Radii of single-point V=0.3 uniform disk fits as in
 Figure~\ref{fig905Diams} for 
 the K bandpass, which is moderately contaminated by the H$_2$O molecule.}
 \label{figKDiams}
\end{figure}

\begin{figure}
 \includegraphics{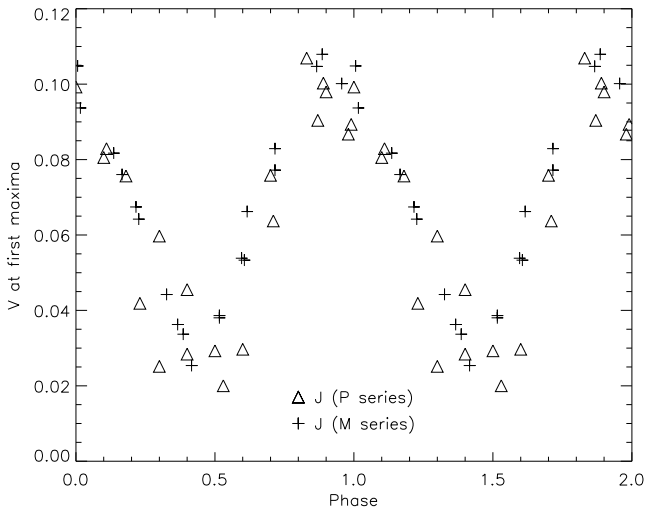}
 \caption{Height of the first maxima in the visibility curve for the
 J band for both the P and M series, which is a measure of 
 the limb-darkening in this bandpass.}
 \label{figJLD}
\end{figure}

In this section, the behaviour of simulated high angular resolution 
imaging data products based on the P and M model series is investigated.
Although, in the general case, the full model center-to-limb variation
(CLV) will be required to interpret the output of a high resolution
measurement, here we choose a simple but instructive metric.
A uniform disk (UD) profile has been fitted to the computed model 
visibility function at a single spatial frequency where the visibility
is first less than 0.3.
In the case where the CLV strongly resembles a UD, it doesn't matter 
where the fit is made, and in the case where the CLV departs from a
UD, the fit at 0.3 gives a fairly robust estimator of overall ``size''
without strong influence from the exact form of the profile. 
For complicated CLVs, different fitting profiles and spatial 
frequency ranges will yield varying results (see ISW04 for a 
discussion of this relating to the P series).

Figure~\ref{figFitDiams} shows single-point UD radii from fits to 
the M model series.
The cycle-to-cyle variation seen
here is much less than in the P series as shown in the eqivalent figure
of ISW04 (Figure~5 in that paper). 
The bandpass differing most between the two series appears to be the 
L* bandpass (centred on 3.799~$\mu$m, width 0.6~$\mu$m) which has
much smaller apparent radii in the M series than in the P
series. Note, however, that though this near-continuum bandpass avoids the 
strong-H$_2$O regime at the short-wavelength end of the conventional
L bandpass, it is still substantially affected by fairly strong water
absorption partly originating in relatively high atmospheric layers which have
significant modelling uncertainties (see Section 2). 

In order to make a closer comparison of key bandpasses, we show fit
radii for both model series for both cycles folded in together
in Figures~\ref{fig905Diams} (905~nm bandpass, centred at 0.905~$\mu$m,
width 0.05~$\mu$m) and \ref{figKDiams} (K bandpass). These bandpasses
were chosen as they are moderately contaminated by the
important molecules TiO and H$_2$O, but are not sensitive to the
low-temperature high layers in the models. 
For a more detailed discussion of molecular contamination including CLVs
and optical depth radii, see JS02.
It is clear that the extreme cycle of the P series 
shows significantly larger apparent
radii than either cycle of the M series, but if one were to pick
a cycle at random, there is no systematic difference in these
observables between the two model series. These figures also display
the $\tau_{1.04}$=1 optical-depth radii $R_{1.04}$ in the narrow 
1.04~$\mu$m continuum bandpass for the two series, which shows a 
variation of radius with phase that is only modestly different
between the two series. First, we see that the $R_{1.04}$ pulsation 
amplitude is about 20 percent smaller for the more massive M series.
However even in near-continuum bandpasses, this difference in 
$R_{1.04}$ is partly masked by small amounts of molecular contamination
in interferometric diameter measurements.

It is important to note also that the parent-star radius $R_p$
used for normalization in the figures is 8 percent larger for the M models.
If two stars have the same core mass and, hence, the same luminosity but
have different envelope mass, simultanous measurements of the diameter in a
weak-contamination bandpass and of multi-bandpass fluxes (for deriving the
bolometric flux), should reveal a systematic difference in the effective
temperature defined (e.g. $T_{1.04}$) in terms of the diameter in that
bandpass. The higher-mass star is systematically cooler, and differences are
phase-dependent (see $T_{1.04}$ entries in Table 1 of ISW04 and of this paper).
Investigation of both the amplitude effect and the temperature effect requires 
high-precision interferometry and successive observations in a close phase 
sequence, covering a considerable phase interval and possibly several cycles,
and has to be done as a differential comparison between two or more stars.
A promising way to accomplish this would be closely-spaced ($\sim$weekly)
near-continuum interferometry with simultaneous photometry catching 
pre-maximum phases of maximal compression and almost UD-like intensity
distribution. A robust and precise measurement here should yield a maximum
effective temperature which is little affected by upper-atmosphere
molecular contamination and is systematically higher in the lower-mass
P models than in the M models (see Table 1 in this paper and in ISW04).

The standard bandpass that shows minimum contamination by any molecule
is the J bandpass. Although the values for the fit radius are very
similar between the two series and are close to $R_{1.04}$, one might
expect the different
density profiles of the two model series to produce different shapes
of their visibility curves. A means to test this hypothesis is to
measure the height of the first maxima in the curve of visibility
modulus versus baseline, which gives an estimate of limb-darkening. 
These values are plotted for both cycles of both series in
Figure~\ref{figJLD}. This measurement would be 0.13 for a uniform disk,
and 0.0 for a Gaussian CLV. It is clear that
there is no systematic difference between the two series in this
observable, implying similar brightness profiles.

As with the lightcurve analysis in the previous section, the strongest
discriminant in the simulated observables from the P and the M model
series appears to be the cycle-to-cycle variation.
The lower-mass P models are certainly seen to produce more erratic 
behavior. It is debatable whether the present models can be
trusted to the point of delivering a quantitative mass diagnostic 
based on these observed effects (see Section 2). However, differential
studies of different stars may shed some light on relative
differences in mass.

\section{Discussion}

Changes of the structure of deeper atmospheric layers (in which the
continuum is formed and in which the contamination of standard 
near-continuum bandpasses is generated) are very modest when the
envelope mass of the Mira is increased by a factor of 1.5 from 
0.4 to 0.6 $M_{\odot}$. The resulting light curves and disk brightness
distributions in these bandpasses are very similar for the here investigated
P (envelope 0.4 $M_{\odot}$) and M (0.6) model series.
A more significant mass dependence of the CLV one might suspect from the
studies of HSW98 and JS02, is feigned by the strong phase dependence of the
CLV shape. This phase dependence can only be recognized on the basis of a 
closely sampled time sequence. In fact, considering the fundamental 
uncertainties in phase assignment, sound interpretation of an interferometric 
diameter measurement at an isolated single phase may be extremely 
difficult. Significant mass effects {\em do} occur in the outer atmospheric
layers which, however, are not readily accessible to model analysis owing
to serious problems of modelling these layers.

The most conspicuous difference between the 1 $M_{\odot}$ P series and the
1.2 $M_{\odot}$ M series is the pronounced cycle-to-cycle variation of the
P model series that is not found in the M series. Though quantitative 
details may be questioned in the present stage of modelling (Section 2),
it is obvious from, for instance, light curves (e.g. Whitelock et al. 2000), 
TiO+VO colour index observations (e.g. Spinrad \& Wing 1969),
radial-velocity measurements (e.g. Joy 1954) and a few 
measurements of time-dependent diameters (e.g. Tuthill et al. 1995; 
Burns et al. 1998; Thompson et al. 2002) that noticeable cycle-to-cycle 
variations exist in Mira variables. From the light curves shown in 
Figures 3 and 5 (where the modelled curves in the L bandpass have 
to be considered with great caution, see Section 3), we suspect that the 
differences between the here selected two rather ``extreme" cycles of 
the P series are larger than the cycle-to-cycle differences found 
in $o$~Cet whereas the M series appears closer to a real star. Note that
the two prototype Miras $o$~Cet and R~Leo that have very similar 
periods and spectral types ($o$~Cet M5e-M9e, R~Leo M6e-M9.5e, Whitelock
et al. 2000) and are not known to show obvious spectral anomalies, have
very different light curves in the K and L bandpasses. The amplitudes
of R~Leo are conspicuously small in K and L and cannot be predicted by any
presently available model. We cannot exclude but seriously doubt that more 
pronounced mass variations would lead to a better model representation 
of R~Leo.

For given period and luminosity, metallicity is the second fundamental
parameter of a Mira variable besides mass. Metallicity is known to influence 
observable properties significantly (cf. Scholz \& Wood 2004) and effects 
of changing the chemical composition are under investigation, but no 
quantitative results are available yet. Another source of differences
between seemingly similar Miras are patchy structures of the atmosphere
and other deviations from spherical symmetry, which have been observed in
numerous stars including $o$~Cet and R~Leo (see the literature listed in
Scholz 2003) but are outside the scope of the present spherical models. The
conspicuous decrease of R~Leo's amplitude (compared to $o$~Cet and the models)
with increasing wavelength (J, K, L) could be due to
either an effect of metallicity or non-spherical symmetry, or
possibly a modest difference in luminosity between the two stars 
within the period-luminosity relationship scatter. It
might also indicate the importance of a
high- to circumstellar-layer phenomenon like, e.g., dust absorption that 
may depend sensitively upon the stellar mass (or metallicity) but is not 
properly described by our simple models. Note, however, that the
simple models of Schuller et al. (2004) predict that dust
optical-depths in both absorption and scattering are too small in
R~Leo to cause significant effects in these bandpasses.

\section{Conclusions}

An extensive study contrasting the observable properties of M-type Mira 
variables based on models of significantly different mass (1 $M_\odot$ and 
1.2 $M_\odot$ or envelope masses of 0.4 $M_\odot$ and 0.6 $M_\odot$) has 
been undertaken. 
No unequivocal signatures discriminating between these was found within the 
parameter space explored, covering observations of the lightcurves, CLV or 
apparent size behavior in selected filter bandpasses over the pulsation 
cycle. 
Three observational strategies, reliant on more precise and/or extensive
measurement, showed promise in distinguishing high from low mass
stars\footnote{Predictions of existing and additional models are
  available upon request to anyone interested for specific
  observational programs.}.
The first of these would entail close observation covering a number of
pulsation cycles, with the lower mass objects exhibiting greater  
cycle-to-cycle variations of spectrophotometric and interferometric 
properties.
Other possible approaches are to search for small differences in
the pulsation amplitude or in effective temperatures through careful 
and relatively precise measurement of apparent diameter in an (almost) 
uncontaminated bandpass.

\section*{Acknowledgments}
This research was supported by the Australian Research Council and
the Deutsche Forschungsgemeinschaft within the linkage project ``Red Giants''.

\label{lastpage}

\end{document}